\documentstyle[preprint,aps]{revtex}

\begin{document}
\title{{\bf Quantum Teleportation in a Solid State System}}
\author{John H. Reina and Neil F. Johnson}
\address{Physics Department, Clarendon Laboratory, Oxford University,\\
Oxford OX1 3PU, U.K.}
\maketitle
\date{\today }

\begin{abstract}
We propose a practical solid-state system capable of demonstrating quantum
teleportation. The set-up exploits recent advances in the optical control of
excitons in coupled quantum dots, in order to produce maximally-entangled
Bell and Greenberger-Horne-Zeilinger (GHZ) states. Only two unitary
transformations are then required: a quantum Controlled-Not gate and a
Hadamard gate. The laser pulses necessary to generate the
maximally-entangled states, and the corresponding unitary transformations,
are given explicitly.
\end{abstract}

\newpage

Since the original idea of quantum teleportation considered in 1993 by
Bennett {\it et al}. \cite{Bennett1}, great efforts have been made to
realize the physical implementation of teleportation devices \cite
{Teleportation}. The general scheme of teleportation \cite{Bennett1}, which
is based on Einstein-Podolsky-Rosen (EPR) pairs \cite{Einstein} and Bell
measurements \cite{Bell} using classical and purely nonclassical
correlations, enables the transportation of an arbitrary quantum state from
one location to another without knowledge or movement of the state itself
through space. This process has been explored from various points of view 
\cite{Teleportation}; however none of the experimental set-ups to date have
considered a solid-state approach, despite the recent advances in
semiconductor nanostructure fabrication and measurement \cite
{Bonadeo1,Bonadeo2,ChavezPirson,Johnson}. Reference \cite{Bonadeo1}, for
example, demonstrates the remarkable degree of control which is now possible
over quantum states of individual quantum dots (QDs) using ultra-fast
spectroscopy. The possibility therefore exists to use optically-driven QDs
as ``quantum memory'' elements in quantum computation operations, via a
precise and controlled excitation of the system. In this Letter we propose a
practical scheme for quantum teleportation which exploits currently
available ultrafast spectroscopy techniques in order to prepare and
manipulate entangled states of excitons in coupled QDs. To our knowledge,
this is the first practical proposal for demonstrating quantum teleportation
in a solid state system.

In order to implement the quantum operations for the description of the
teleportation scheme proposed here, we employ two elements: the quantum
controlled-not\ gate ({\it CN}$\;$gate){\it ,} and the Hadamard
transformation. In the orthonormal computation basis of single qubits $%
\left\{ \left| 0\right\rangle ,\left| 1\right\rangle \right\} $, the $CN$
gate acts on two qubits $\left| \varphi _{i}\right\rangle $ and $\left|
\varphi _{j}\right\rangle $ simultaneously as follows: $CN_{ij}(\left|
\varphi _{i}\right\rangle \left| \varphi _{j}\right\rangle )\mapsto \left|
\varphi _{i}\right\rangle \left| \varphi _{i}\oplus \varphi
_{j}\right\rangle $.$\;$Here $\oplus $ denotes addition\ modulo\ 2. The
indices $i$ and $j$ refer to the control bit and the target bit
respectively. The Hadamard transformation $H\;$acts only on single qubits by
performing the rotations $H(\left| 0\right\rangle )\mapsto \frac{1}{\sqrt{2}}%
\left( \left| 0\right\rangle +\left| 1\right\rangle \right) \;$and $H(\left|
1\right\rangle )\mapsto \frac{1}{\sqrt{2}}\left( \left| 0\right\rangle
-\left| 1\right\rangle \right) $. We also introduce a pure state $\left|
\Psi \right\rangle $ in this Hilbert space given by $\left| \Psi
\right\rangle =\alpha \left| 0\right\rangle +\beta \left| 1\right\rangle $
with\ $\left| \alpha \right| ^{2}+\left| \beta \right| ^{2}=1$, where $%
\alpha \;$and $\beta \;$are $%
\mathop{\rm complex}%
$\ numbers. As discussed later, $\left| 0\right\rangle \;$represents the
vacuum state for excitons while $\left| 1\right\rangle \;$represents a
single exciton.

Figure 1 shows our general computational approach which is inspired by the
work of Brassard {\it et al.} \cite{Brassard}. As usual, we refer to two
parties, Alice and Bob. Alice wants to teleport an arbitrary, unknown qubit
state $\left| \Psi \right\rangle $ to Bob. Figure 2 shows the specific
realization we are proposing using optically controlled quantum dots with QD 
$a$ initially containing $\left| \Psi \right\rangle $. Alice prepares two
qubits (QDs $b$ and $c$) in the state $\left| 0\right\rangle $ and then
gives the state $\left| \Psi 00\right\rangle \;$as the {\it input} to the
system. By performing the series of transformations shown in Fig 1(a), Bob
receives as the {\it output} of the circuit the entangled state $%
{\textstyle {1 \over \sqrt{2}}}%
(\left| 0\right\rangle +\left| 1\right\rangle )_{a}%
{\textstyle {1 \over \sqrt{2}}}%
(\left| 0\right\rangle +\left| 1\right\rangle )_{b}\left| \Psi \right\rangle
_{c}\;$(Fig. 2c). In Fig. 1(b) we extend the analysis of the teleportation
process to the case of a four qubit quantum circuit, which can be realized
by four coupled QDs. As before, Alice wants to teleport the state $\left|
\Psi \right\rangle _{a}$ to Bob. She prepares three qubits in the state $%
\left| 0\right\rangle $ (QDs $b,c$ and $d$) and gives the state $\left| \Psi
000\right\rangle $\ as the input to the system. From Fig. 1(b) it is clear
that the function of the first three operations performed by Alice is to
obtain the maximally entangled GHZ state $%
{\textstyle {1 \over \sqrt{2}}}%
(\left| 000\right\rangle +\left| 111\right\rangle )$.$\;$The next two
operations realized by Alice (before the arrow in Fig 1(b)) leave the system
in the state 
\begin{equation}
\begin{array}{c}
{\textstyle {1 \over 2}}%
\left\{ \left| 000\right\rangle (\alpha \left| 0\right\rangle +\beta \left|
1\right\rangle )+\left| 011\right\rangle (\beta \left| 0\right\rangle
+\alpha \left| 1\right\rangle )+\right.  \\ 
\left. +\left| 100\right\rangle (\alpha \left| 0\right\rangle -\beta \left|
1\right\rangle )+\left| 111\right\rangle (-\beta \left| 0\right\rangle
+\alpha \left| 1\right\rangle )\right\} \text{.}
\end{array}
\end{equation}
Performing the operations shown after the arrow in Fig. 1(b), Bob gets as
the output of the circuit the entangled state $%
{\textstyle {1 \over \sqrt{2}}}%
(\left| 0\right\rangle +\left| 1\right\rangle )_{a}%
{\textstyle {1 \over \sqrt{2}}}%
\left( \left| 00\right\rangle +\left| 11\right\rangle \right) _{b,c}\left|
\Psi \right\rangle _{d}$.$\;$Hence the state $\left| \Psi \right\rangle $
was teleported from dot $a$ to dot $d$ in the system. In order to describe
in detail how this circuit may be implemented, we need to perform the
following steps: Alice prepares three qubits in the state $\left|
0\right\rangle $, and then sends the first two of them through the two first
gates between QDs $b$ and $c$, as shown in Fig. 1(b). She keeps the
information stored in QD $c$, namely $\gamma $, in her quantum memory and
sends qubit $\beta \;$to Bob. In the next step she pushes the other qubit
which is in state $\left| 0\right\rangle $ together with $\gamma $ to the
third gate; after this operation, she keeps the last qubit of the system, $%
\delta $. Alice then receives from QD $a$ the qubit $\left| \Psi
\right\rangle $ which she wants to teleport to Bob. To achieve this, she
removes the $\delta $ qubit from her quantum memory and sends this, together
with qubit $\left| \Psi \right\rangle $,$\;$to the next two gates of the
circuit. She then performs a measurement of the output between QDs $a$ and $c
$ \cite{Note1} (at the arrow in Fig. 1(b)) in order to turn the result into
two classical bits $\Lambda $ and $\Gamma \;$respectively. To finish the
teleportation process, Alice needs to communicate $\Lambda $ and $\Gamma $
to Bob via a classical communication channel. Hence after the arrow, Bob
receives the classical information and creates the quantum states $\left|
\Lambda \right\rangle $ and $\left| \Gamma \right\rangle $. Next, he removes
the qubits $\beta \;$and$\;\delta \;$from his quantum memory and sends the
four qubits to his part of the circuit. At this point teleportation is
complete since Bob receives at his output the state $\left| \Psi
\right\rangle $ on dot $d$. Hence our quantum teleportation circuit (QTC)
transforms the input$\;\left| \Psi \right\rangle _{a}\left| 0\right\rangle
_{b}\left| 0\right\rangle _{c}\left| 0\right\rangle _{d}\;$into the$\;$output%
$\;{{\frac{1}{\sqrt{2}}}}(\left| 0\right\rangle +\left| 1\right\rangle
)_{a}\left| Bell\right\rangle _{b,c}\left| \Psi \right\rangle _{d}$.$\;$

Interestingly, the above teleportation process can be extended to an $n-$QTC
using the Schr\"{o}dinger's cat state, as shown in Fig. 1(c). Again, the
goal is to teleport the state $\left| \Psi \right\rangle _{a}$ from Alice to
Bob. She prepares $n-1$ qubits in the state $\left| 0\right\rangle $ (QDs$%
\;b,...,$ $n$) and hence gives the state $\left| \Psi 00...0\right\rangle \;$%
as the input$.$ After the first $n-1$ operations (Fig 1(c)) she obtains the
Schr\"{o}dinger's cat state $%
{\textstyle {1 \over \sqrt{2}}}%
\left( \left| 00...0\right\rangle +\left| 11...1\right\rangle \right)
_{b,...,n}\;$which, followed by the last two exclusive-ors operations before
the arrow, leaves the system in the following state of $n$ qubits 
\begin{equation}
\begin{array}{c}
{\textstyle {1 \over 2}}%
\left\{ \left| 00...0\right\rangle (\alpha \left| 0\right\rangle +\beta
\left| 1\right\rangle )+\left| 011...1\right\rangle (\beta \left|
0\right\rangle +\alpha \left| 1\right\rangle )+\right.  \\ 
\left. +\left| 100...0\right\rangle (\alpha \left| 0\right\rangle -\beta
\left| 1\right\rangle )+\left| 11...1\right\rangle (-\beta \left|
0\right\rangle +\alpha \left| 1\right\rangle )\right\} \text{.}
\end{array}
\end{equation}
The procedure to realize the circuit of Fig. 1(c) follows directly from the
description given for Fig. 1(b). In the case of Fig 1(c), the measurement
performed by Alice at the end of her part of the circuit, must be realized
between the $a-$th and the $(n-1)$th QDs \cite{Note1} in order to turn the
result into two classical bits $\Theta $ and $\Upsilon \;$respectively.
Hence Alice communicates these bits to Bob via a classical channel and,
after the arrow, Bob receives the classical information and creates the
quantum states $\left| \Theta \right\rangle $ and $\left| \Upsilon
\right\rangle $. Next, he removes from his quantum memory the other $n-2\;$%
qubits, thereby ultimately obtaining $\left| \Psi \right\rangle $ on the $n$%
th$-$dot. In this way, the QTC presented here transforms the input state$%
\;\left| \Psi \right\rangle _{a}\left| 0\right\rangle _{b}\left|
0\right\rangle _{c}...\left| 0\right\rangle _{n}\;$into the$\;$output$\;{{%
{\textstyle {1 \over \sqrt{2}}}%
}}(\left| 0\right\rangle +\left| 1\right\rangle )_{a}{{\frac{1}{\sqrt{2}}(}}%
\left| 00...0\right\rangle +\left| 11...1\right\rangle )_{b,...,n-1}\left|
\Psi \right\rangle _{n}$.$\;$From Figs. 1(b) and 1(c) we note that the final
stage of the QTC may be used as a subroutine in larger quantum computations
or for quantum communication; specifically this is because we are recovering
at the output a $2,3,...,\;$or ($n-2$)$-${\it maximally} entangled state 
\cite{Note2}. We also note that the structure of Bob's part of the circuit
is the same for all the circuits in Fig. 1. This is because Bob's function
in the QTC is to realize the ``appropriate rotations'' over the general
state given in Eq. (2).$\;$It is interesting to note that if Bob, instead of
performing the operations after the arrow, chooses one of such appropriate
unitary transformations \cite{Note3} to apply to the $n-$th qubit after
receiving the classical bits from Alice's measurement, then he does not need
to perform his part of the circuit. For this reason only two unitary
exclusive-ors transformations are needed in order to teleport the state $%
\left| \Psi \right\rangle $. However, from the point of view of our
implementation and, more generally, for quantum computer algorithms, it is
better to undertake the complete process shown in the QTC than to choose
such a special rotation. Although our goal is the practical realization
using what is at the limit of current optoelectronics technology, we note
that the circuits of Fig. 1 {\it are not restricted} to QD systems: they can
be applied to any system where the task of entangled-state preparation has
been achieved.

In order to describe the physical implementation of the quantum circuits
using coupled quantum dots, we exploit the recent experimental results
involving coherent control of excitons in single quantum dots on the
nanometer and femtosecond scales \cite{Bonadeo1,Bonadeo2}. Consider a system
of $N$ identical and equispaced QDs containing no net charge which are
radiated by long-wavelength classical light, as illustrated schematically in
Fig. 2(b) for the case $N=3$. The formation of single excitons within the
individual QDs and their inter-dot hopping can be described by the
Hamiltonian \cite{Quiroga,Reina} 
\begin{equation}
H(t)=\epsilon J_{z}+W(J^{2}-J_{z}^{2})+\xi (t)J_{+}+\xi ^{*}(t)J_{-}\text{,}
\end{equation}
where $J_{+}=%
\mathop{\textstyle \sum }%
\limits_{p=0}^{N}c_{p}^{\dagger {}}h_{p}^{\dagger
{}},\;J_{-}=\sum\limits_{p=0}^{N}h_{p}c_{p},\;$and $J_{z}=%
{\textstyle {1 \over 2}}%
\sum\limits_{p=0}^{N}\left( c_{p}^{\dagger {}}c_{p}-h_{p}h_{p}^{\dagger
{}}\right) $.$\;$Here $c_{p}^{\dagger {}}\;(h_{p}^{\dagger {}})\;$is the
electron (hole) creation operator in the $p$th QD; $\epsilon $ represents
the QD band gap, $W$ is the interdot interaction parameter, $\xi (t)\;$the
laser pulse shape, while the quasi-spin $J-$operators satisfy the usual
commutation relations $\left[ J_{z},J_{\pm }\right] =\pm \hbar J_{\pm
},\;\left[ J_{+},J_{-}\right] =2\hbar J_{z}\;$and$%
\;[J^{2},J_{+}]=[J^{2},J_{-}]=[J^{2},J_{z}]=0$.$\;$By solving the eigenvalue
problem associated with the time-dependent Hamiltonian (3), we have shown 
\cite{Reina} for several different values of the phase $\phi $, that\ Bell
and GHZ states of the form $%
{\textstyle {1 \over \sqrt{2}}}%
\left( \left| 00\right\rangle +e^{i\phi }\left| 11\right\rangle \right) $ and%
$\;%
{\textstyle {1 \over \sqrt{2}}}%
\left( \left| 000\right\rangle +e^{i\phi }\left| 111\right\rangle \right) \;$%
can be prepared in systems comprising two and three coupled quantum dots,
respectively. The practical requirements are realizable in present
experiments employing both ultrafast \cite{Bonadeo1,Bonadeo2} and near-field
optical spectroscopy \cite{ChavezPirson} of quantum dots. In Figure 3 we
present the generation of $\phi -$pulses which lead to the implementation of
our QTC. As mentioned previously, $\left| 0\right\rangle \;$represents the
vacuum for excitons while $\left| 1\right\rangle \;$denotes a single-exciton
state. For the practical proposal of Fig. 1(a), we require 3 equidistant QDs
which initially must be prepared in the state $\left| \Psi \right\rangle
_{a}\left| 0\right\rangle _{b}\left| 0\right\rangle _{c}$. As shown in Fig.
2(a), one of these (QD $a$) contains the quantum state $\left| \Psi
\right\rangle _{a}$ that we wish to teleport, while the other two (QDs $b$
and $c$) are initialized in the state $\left| 00\right\rangle _{bc}-\;$this
latter state is easy to achieve since it is the ground state. Following this
initialization,$\;$we illuminate QDs $b$ and $c$ with a radiation pulse of
frequency $\omega \;$given by $\xi (t)=A\exp (-i\omega t)\;$(see Fig. 2(b));
here $A$\ includes the electron-photon coupling and the electric field
strength, and the time-dependence of $\xi $ gives the pulse shape. As an
example we consider the case of ZnSe-based QDs. The band gap $\epsilon
=2.8\;e$V, hence the resonance optical frequency $\omega =4.3\times 10^{15}\;
$s$^{-1}$.$\;$In units of $\epsilon $,$\;W=0.1\;$and$\;A=%
{\textstyle {1 \over 25}}%
$.$\;$For a $0$ or $2\pi -$pulse, the density of probability for finding the
QDs $b$ and $c\;$in the Bell state\thinspace $%
{\textstyle {1 \over \sqrt{2}}}%
\left( \left| 00\right\rangle +\left| 11\right\rangle \right) $ shows that a
time $\tau _{_{Bell}}=7.7\times 10^{-15}\;$s is required (see Fig. 3(a)).$\;$%
This time $\tau _{_{Bell}}\;$hence corresponds to the realization of the
first two gates of the circuit in Fig. 1(a), i.e. the Hadamard
transformation (or$\;%
{\textstyle {\pi  \over 4}}%
-$rotation) over QD $b$ followed by the {\it CN} gate between QDs $b$ and $c$%
. After this, the information in qubit $c$ is sent to Bob and Alice keeps in
her memory the state of QD $b\;$(Fig. 1(a))$.$ Next, we need to perform a 
{\it CN} operation between QDs $a\;$and $b$ and, following that, a Hadamard
transform over the QD $a$: this procedure then leaves the system in the
entangled state 
\begin{equation}
{\textstyle {1 \over 2}}%
\left\{ \left| 00\right\rangle (\alpha \left| 0\right\rangle +\beta \left|
1\right\rangle )+\left| 01\right\rangle (\beta \left| 0\right\rangle +\alpha
\left| 1\right\rangle )+\left| 10\right\rangle (\alpha \left| 0\right\rangle
-\beta \left| 1\right\rangle )+\left| 11\right\rangle (-\beta \left|
0\right\rangle +\alpha \left| 1\right\rangle )\right\} \text{.}
\end{equation}
The last step is possible in practice using the masking technique of
exciting and detecting the dot luminiscence through micrometer-sized
apertures in an aluminum mask \cite{Bonadeo1}; this allows for selection of
a few QDs within the broad distribution of dots in the sample. Combining
spatial and spectral resolutions, it therefore becomes possible to excite
and probe only one individual QD{\it \ }with the corresponding dephasing
time being $\tau _{d}=4\times 10^{-11}\;$s$\;$\cite{Bonadeo1}.$\;$Hence we
have the possibility of coherent optical control of the quantum state of a
single dot. Furthermore, this mechanism can be extended to include more than
one excited state: since $%
{\displaystyle {\tau _{_{Bell}} \over \tau _{d}}}%
\simeq 1.8\times 10^{-4}$,$\;$several thousand unitary operations can in
principle be performed in this system before the excited state of the QD
decoheres. This fact together with the experimental feasibility of applying
the required sequence of laser pulses on the femtosecond time-scale \cite
{Erland} leads us to conclude that we do not need to worry unduly about
decoherence ocurring whilst performing the other four unitary operations
that Bob needs in order to obtain the final state ${{%
{\textstyle {1 \over \sqrt{2}}}%
}}(\left| 0\right\rangle +\left| 1\right\rangle )_{a}{{\frac{1}{\sqrt{2}}(}}%
\left| 0\right\rangle +\left| 1\right\rangle )_{b}\left| \Psi \right\rangle
_{c}$,$\;$thereby completing the teleportation process. In the case of Fig.
3(b), a similar analysis shows that $\tau _{_{GHZ}}=1.3\times 10^{-14}\;$s,\
and hence$\;%
{\displaystyle {\tau _{_{GHZ}} \over \tau _{d}}}%
\simeq 3.3\times 10^{-4}$: this also makes the circuit in Fig. 1(b)
experimentally feasible. Although this discussion refers to ZnSe-based QDs,
other regions of parameter space can be explored by employing semiconductors
of different bandgap $\epsilon $. A more detailed description of the
implementation of the {\it CN} and the Hadamard operations will be discussed
elsewhere \cite{Reina}. Even though the\ structures that we are considering
have a dephasing time of order $10^{-11}\;$s,$\;$QDs with stronger
confinement are expected to have even smaller coupling to phonons giving the
possibility for much longer intrinsic coherence times.

In summary, we have proposed a practical implementation of a quantum
teleportation device, exploiting current levels of optical control in
coupled QDs. Furthermore the analysis suggests that several thousand quantum
computation operations may in principle be performed before decoherence
takes place.

The authors thank L. Quiroga, J. Erland, D.J.T. Leonard and S.C. Benjamin
for helpful discussions. J.H.R. thanks the financial support of COLCIENCIAS
(Colombia) and gratefully acknowledges H. Steers for continuous
encouragements.

\begin{center}
\newpage

{\large FIGURE CAPTIONS}
\end{center}

FIG. 1. Circuit schemes to teleport an unknown quantum state from Alice to
Bob using an arrangement of (a) 3, (b) 4 and (c) $n$ qubits (coupled quantum
dots). The methods employ (a) Bell, (b) GHZ, and (c) Schr\"{o}dinger's cat
states respectively.\ The operator subscripts denote the qubits being
addressed. For simplicity the output is not shown in (c).\bigskip

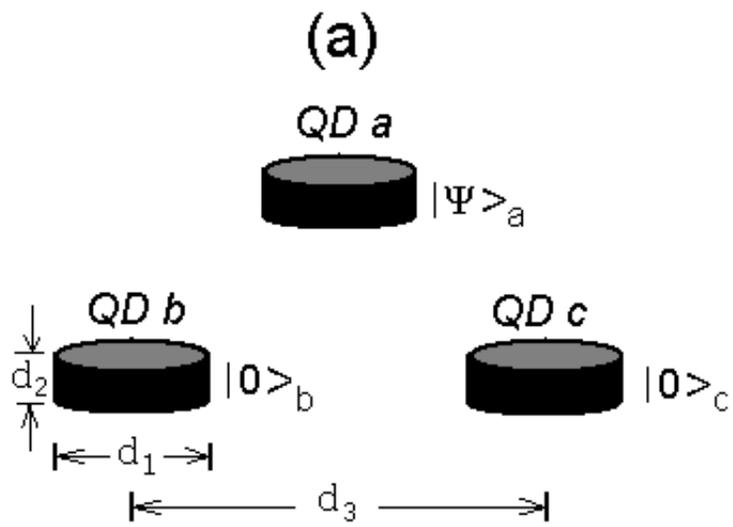
FIG. 2. Practical implementation of teleportation using optically-driven
coupled quantum dots. (a) Initial state of the system. (b) Intermediate
step: radiating the system with the pulse $\xi (t)$. (c) Final state.
Typical values for the dots are diameter $d_{1}=30\;n$m, thickness $%
d_{2}=3\;n$m$\;$and separation $d_{3}=100\;n$m.\bigskip

FIG. 3. Generation of (a) Bell and (b) GHZ states. These pulses correspond
to the realization of the Hadamard gate followed by the quantum {\it CN}
gates (see Figs. 1(a) and 1(b)). $\epsilon =2.8\;e$V, $W=0.1$,$\;\phi =0\;$%
and$\;A=%
{\textstyle {1 \over 25}}%
$. Here $\psi (t)\;$denotes the total wavefunction of the system in the
laboratory frame at time $t$.

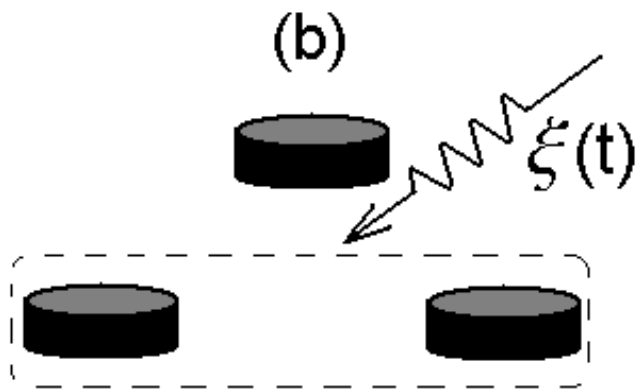
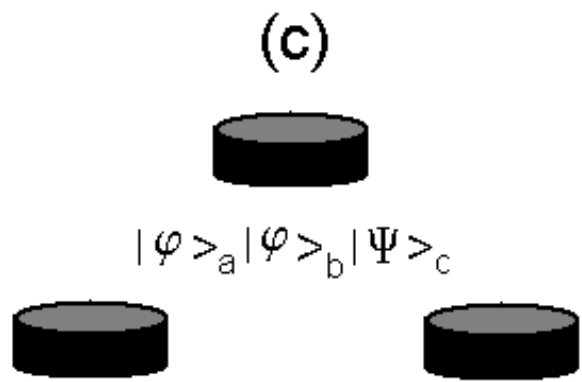
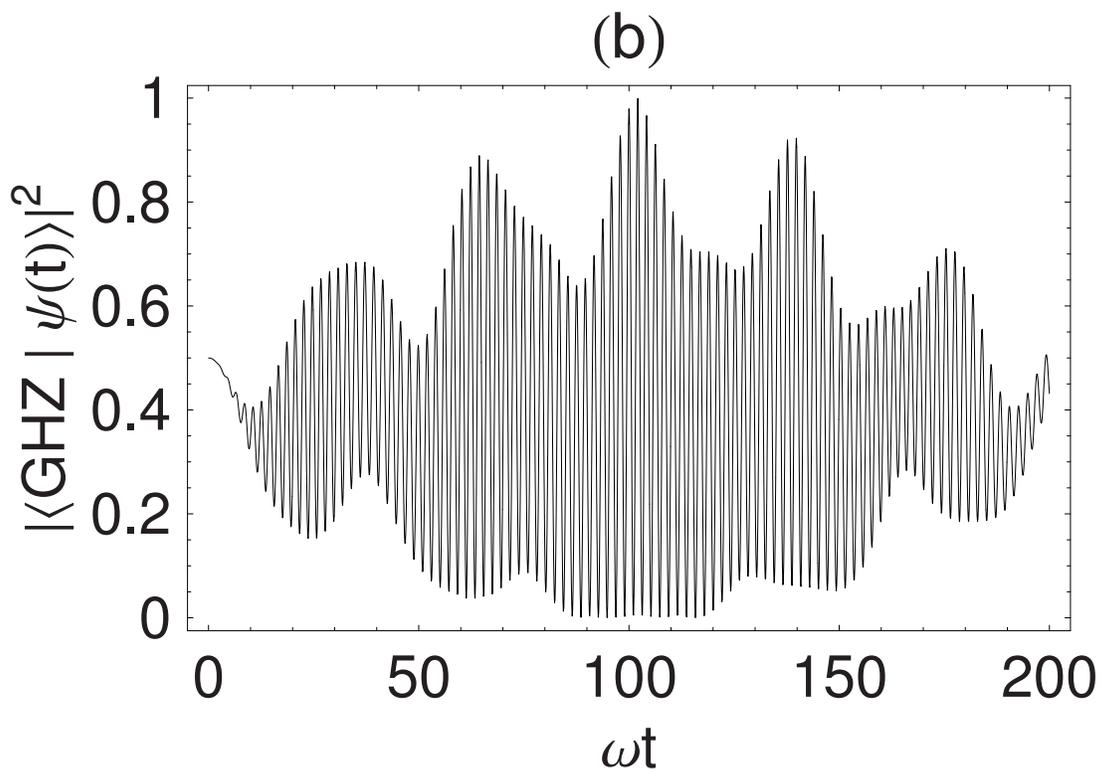
\end{document}